\documentstyle[12pt]{article}
\newcommand{\ti}[1]{\mbox{\tiny{#1}}}

\newcommand{\rmsub}[2]{#1_{\rm #2}}

\def\lsim{~\rlap{$<$}{\lower 1.0ex \hbox{$\sim$}}}
\def\gsim{~\rlap{$>$}{\lower 1.0ex \hbox{$\sim$}}}
\newcommand{\dns}{\mbox{$\rmsub{D}{n} - \sigma \:$}}

\newcommand{\heading}[1]{\vspace*{15mm}
{\Large\begin{center} {\bf{#1}} \end{center}}}
\newcommand{\authors}[3]{\vspace{5mm}
\begin{center}
{\normalsize \rm #1}\\    
\vspace{3mm} {\normalsize \it #2}\\    
\vspace{2mm} {\normalsize \it #3}\\    
\vspace{1.5cm} \end{center} }
\def\lsim{~\rlap{$<$}{\lower 1.0ex\hbox{$\sim$}}}
\def\gsim{~\rlap{$>$}{\lower 1.0ex\hbox{$\sim$}}}
\begin{document}
\heading{Optimal Galaxy Distance Estimators}

\authors{M.A. Hendry$^{1}$ and J.F.L. Simmons$^{2}$}
        {$^{1}$ Astronomy Centre, University of Sussex, Brighton BN1 9QH, UK.}
        {$^{2}$ Department of Physics and Astronomy,
         University of Glasgow, Glasgow G12 8QQ, UK.}

\begin{abstract}
The statistical properties of galaxy distance estimators
corresponding to the Tully-Fisher and \dns relations are
studied, and a rigorous framework for identifying and removing the
effects of Malmquist bias due to observational selection is developed.
The prescription of Schechter (1980) for defining unbiased distance
estimators is verified and extended to more general -- and more
realistic -- cases. Finally, the derivation of 'optimal' unbiased
estimators of minimum dispersion, by utilising information from
additional suitably correlated observables, is discussed and the
results applied to a calibrating sample from the Fornax cluster,
as used in the Mathewson spiral galaxy redshift survey.
The optimal distance estimator derived from apparent magnitude, diameter and
21cm line width has an intrinsic scatter which is 25 \% smaller
than that of the Tully-Fisher relation for this calibrating sample.
\end{abstract}

\clearpage
\newpage

\section{{\sc{introduction}}}
\label{sec:intro}

In recent years the analysis of redshift surveys of galaxies has made a
significant contribution to our emerging understanding of the formation and
evolution of large scale structure in the universe. A crucial element in this
analysis is the accurate estimation of galaxy distances, and an important
feature of many recent surveys has been the availability of redshift
independent distance indicators which allow one to determine directly an
estimate of the radial peculiar velocity of each galaxy in the survey. By far
the most prevalent examples of such distance indicators are the Tully-Fisher
and \dns relations. These have been used by a number of authors in
attempts to reconstruct, from various redshift surveys, the full 3-dimensional
peculiar velocity and density contrast fields (c.f. \cite{lynden88},
\cite{bertetal90}, \cite{potiras93}, \cite{mouldetal93}). This work has been
at the forefront of a mounting body of evidence in support of galaxy
clustering and coherent streaming motions on scales of the order of 100 Mpc;
evidence which, nevertheless, has attracted considerable controversy in the
literature -- not least because of the difficulties which it presents for
currently popular theories of structure formation. Much of this
debate has focussed upon the statistical properties of the Tully-Fisher and
\dns relations, and the extent to which detections of galaxy streaming
might be a statistical artefact of the distance indicators.

The aim of this paper is to address and clarify several statistical issues
relating to the use of redshift independent distance indicators, particularly
with respect to the systematic biases which arise in surveys subject
to observational selection. These systematic effects have been referred to
generically in the literature as `Malmquist bias', although there exists a lack
of consensus as to precisely what is meant by this term -- and consequently
some disagreement over how one should best deal with its effects in analyses
of galaxy redshift surveys.
In this paper we identify Malmquist bias and examine its effects on
redshift independent distance indicators by following the statistical
formalism which we adopted previously in this context in
\cite{hendrysimmons90} (hereafter HS). In particular we examine in
what circumstances Malmquist bias may be eliminated completely from redshift
independent distance indicators, thus defining what one might regard as an
`optimal' galaxy distance estimator.
We will also consider the statistical basis of other approaches to
Malmquist bias which have been adopted in the literature (c.f.
\cite{lynden88}, \cite{landsz92}), and clarify the important differences
between these approaches and the formalism which we adopt here. In a concurrent
paper \cite{newsametal93} we examine in detail the consequences of using
biased distance indicators for reconstructing the large scale velocity and
density fields -- particularly with respect to the {\sc potent} method
\cite{bertetal90}, \cite{potiras93}.

The Tully-Fisher and \dns relations are both derived empirically, by fitting a
power law to the relationship between two intrinsic physical characteristics
of galaxies: the luminosity and the width of the HI 21cm line of spirals in
the case of Tully-Fisher, and the intrinsic diameter and central velocity
dispersion of ellipticals in the case of \dns. Both relations are generally
expressed in terms of log quantities, and are thus fitted to be linear in
form -- e.g. for Tully-Fisher we have an expression of the form:-
\begin{equation}
\label{eq:TFreln}
\centering
  M = a \, \log W \, + \, b
\end{equation}
where $M$ is the absolute magnitude and $W$ the 21cm line width. The constants
$a$ and $b$, the slope and zero-point of the relation, are
determined empirically -- usually with a calibrating sample of reference
galaxies the distances of which have been measured independently.\hspace{-2mm}
\footnote{In some analyses of redshift surveys, c.f. \cite{hanmould90}, the
slope and zero point are fitted simultaneously with the parameters of a
specific velocity field model, using all of the survey galaxies. We will
consider the specific statistical issues raised by this contrasting approach
elsewhere.}
To apply the
relation one simply measures the line width of a given galaxy, and infers from
equation \ref{eq:TFreln} an estimate of its absolute magnitude. This can then
be combined with the galaxy's observed {\em apparent \/} magnitude to obtain an
estimate of its distance.

Finding the `best' values of the constants $a$ and $b$ has been a thorny issue
in the literature for a number of years. The straight line relationship given
by equation \ref{eq:TFreln} is generally fitted by performing a linear
regression on the calibrating sample. The question of which linear regression
is most appropriate is non-trivial, however, particularly when the one's
survey is subject to observational
selection effects -- a fact which has been widely recognised (c.f.
\cite{schechter80}, \cite{teerikorpi84}, \cite{tully88}, \cite{hendry92},
\cite{bicknell92}). We can illustrate this with the following simple
example. Figure (1) represents schematically the typical
scatter of the Tully-Fisher relation, assuming that absolute magnitude and
log line width are random variables whose joint distribution is bivariate
normal. (More precisely, the ellipse shown in Figure (1) is an
isoprobability contour enclosing a given confidence region for magnitude and
log line width). The solid and dotted lines indicate the linear relationships
obtained by regressing line widths on
magnitudes and magnitudes on line widths respectively. Thus the dotted line is
the mean, or {\em expected \/}, value of absolute magnitude conditional upon
log line width. Conversely the solid line is the expected log line width
conditional upon absolute magnitude. Since in practice one wishes to infer
the value of M from the measured line width, the regression of magnitudes on
line widths is generally referred to as the `direct' Tully-Fisher relation,
while regressing line widths on magnitudes is often termed the `inverse'
Tully-Fisher relation. Introducing {\bf P} as a shorthand for log line width
(c.f. \cite{teerikorpi84}, \cite{hendry92}, \cite{bicknell92}), the
following equations define the direct and inverse regression lines for the
bivariate normal case:-
\begin{equation}
\label{eq:TFdir}
\centering
  E({\bf M}|{\bf P}) = \rmsub{M}{\ti{0}} + \rho
  \frac{\rmsub{\sigma}{\ti{M}}}{\rmsub{\sigma}{\ti{P}}}
  ({\bf M} - \rmsub{M}{\ti{0}})
\end{equation}
\begin{equation}
\label{eq:TFinv}
\centering
  E({\bf P}|{\bf M}) = \rmsub{P}{\ti{0}} + \rho
  \frac{\rmsub{\sigma}{\ti{P}}}{\rmsub{\sigma}{\ti{M}}}
  ({\bf P} - \rmsub{P}{\ti{0}})
\end{equation}
where $\rmsub{M}{\ti{0}}$, $\rmsub{P}{\ti{0}}$, $\rmsub{\sigma}{\ti{M}}$,
$\rmsub{\sigma}{\ti{P}}$
and $\rho$ denote the means, dispersions and correlation coefficient of the
bivariate normal distribution of magnitudes and log line widths. Note that we
have also adopted the standard statistical convention of denoting random
variables by bold face characters.

It follows from equations \ref{eq:TFdir} and \ref{eq:TFinv} that both the
direct and inverse regression lines can be used to infer an estimate of the
absolute magnitude of a given galaxy which is a linear function of its
measured log line width, although the constants $a$ and $b$ in equation
\ref{eq:TFreln} will clearly be different in each case. Moreover the
definition of the estimate of M inferred from each regression line is also
subtly different. For the direct regression line the estimate of M is the
mean absolute magnitude at the observed log line width -- i.e.
$E({\bf M}|{\bf P} = \rmsub{P}{\ti{obs}})$.
For the inverse regression line, on the other hand, the estimated absolute
magnitude is the value of ${\bf M}$ such that the mean log line width
conditional upon ${\bf M}$ is equal to its observed value -- i.e.
$E({\bf P}|{\bf M}) = \rmsub{P}{\ti{obs}}$. Consequently -- as indeed is
apparent from their slopes -- the two lines give rise to markedly different
distance estimators.

The situation becomes more complex when we include the effects of
observational selection. Figure (2) shows schematically the
distribution of ${\bf M}$ and ${\bf P}$ for {\em observable \/} galaxies in a
sample subject to a sharp cut-off in absolute magnitude -- as would be the case
for e.g. a more distant cluster observed in an apparent magnitude limited
survey.
We can see that in this case the expected value of ${\bf M}$ conditional on
${\bf P}$ is dramatically different from the direct regression line for the
complete sample: in fact $E({\bf M} | {\bf P})$ is no longer linear in
${\bf P}$ but curves sharply as ${\bf M}$ approaches the magnitude limit.

This means that if one calibrates the Tully-Fisher relation in the nearby
cluster using the direct regression line, and then applies this relation to
estimate the distance of a more distant cluster (or indeed a distant field
galaxy), one will systematically underestimate its distance because the
expected value of ${\bf M}$ given ${\bf P}$ in the more distant
sample is systematically brighter than the value predicted by the direct
regression line. It is essentially this systematic error or {\em bias \/} in
the inferred distance which we identify as `Malmquist bias', although we
will define more rigorously what we mean by the bias of a distance
estimator in section \ref{sec:genlin} below. The bias is precisely analogous
to the effect identified by \cite{malmquist20} in considering the mean
absolute magnitude of observable `standard candles' brighter than some given
apparent magnitude limit. The effect of Malmquist bias
upon the Tully-Fisher relation has been illustrated in a similar manner to
Figures (1) and (2) by a number of different
authors (c.f. \cite{lynden88}, \cite{tully88}, \cite{bicknell92}).

For the case where one's sample is subject {\em only \/} to luminosity
selection \cite{schechter80} recognised that the slope of the inverse
regression line is unchanged, irrespective of the magnitude-completeness of
one's sample. \hspace{-3.5mm}
\footnote{Schechter's original treatment was for the Faber Jackson relation
between luminosity and velocity dispersion for ellipticals, although he noted
that precisely the same principle held for Tully-Fisher.}
In other words this regression line is free from the Malmquist effect, and it
may therefore be used to provide an unbiased galaxy distance estimate.
Although the unbiased property of the `Schechter' inverse regression line has
been generally recognised, its ramifications for estimating galaxy distances
have not -- it would seem to us -- been fully appreciated, and the application
of Schechter's ideas to more realistic situations has not been fully explored.
Such an extension forms the central aim of this paper. We set out to place the
Schechter result on a rigorous statistical footing, following the same
formalism previously developed in HS, in order to confirm its range of
validity, examine the assumptions upon which it depends and consider
to what extent those assumptions may be generalised.

To this end, in section \ref{sec:genlin} we study the properties
of a general distance estimator \hspace{-2.5mm}
\footnote{More correctly, an estimator of {\em log \/} distance -- a point
          to which we return presently}
formed from an arbitrary linear combination of apparent magnitude and log
line width. It is easy to see that whichever linear regression one adopts the
corresponding distance estimator will take this simple linear
form, since one will always infer the absolute magnitude of a given galaxy as
a linear function of log(line width). Our analysis is carried out in the
first instance for the Tully-Fisher case, with the corresponding results for
\dns indicated where appropriate.

\section{{\sc{properties of general linear estimator}}}
\label{sec:genlin}

\subsection{{\sc{what do we mean by malmquist bias?}}}
\label{sec:what}

The approach which we adopt here is a natural extension of the formalism
developed in HS to study the properties of distance estimators which are
functions of only one observable -- apparent magnitude. Before we proceed in
earnest we recall from HS a rigorous definition of what we mean by the {\em
bias \/} of an estimator, and clarify the differences between this approach
and other treatments of Malmquist bias in the literature. The contrasting
approaches can be classed as belonging to one of two categories: `frequentist'
and 'Bayesian'.
These terms reflect the fact that at the heart of the difference between the
two approaches lies the long-standing dichotomy between a Bayesian and
frequentist view of the nature of probability.

The frequentist picture is essentially based on the intuitively familiar
concept that the probability of an event measures the {\em relative
frequency \/} of that event occurring in a large number of repeated experiments
or trials. In the limit as the number of trials tends to infinity a histogram
of relative frequencies tends to the probability density function (pdf) of a
random variable -- in this case our galaxy distance estimator. Crucial to the
frequentist approach is the idea that the true distance of the galaxy in every
trial is a fixed, though of course unknown, parameter -- an `unknown state of
nature' in the usual statistical terminology.

We can state these ideas more rigorously as follows, taking as an
illustration the case of an estimator of log distance since
we have seen in section \ref{sec:intro} that such an estimator arises
naturally from the Tully-Fisher and \dns relations. Estimators of log
distance will be the focus of our analysis for most of this paper, although
similar remarks will clearly also apply to an estimator of distance or any
other parameter.

Suppose that $\rmsub{w}{\ti{0}}$ is the true log distance of a given galaxy.
Let ${\bf \hat{w}}$ denote an estimator of $\rmsub{w}{\ti{0}}$.
(Following the standard convention we denote an estimator of a parameter by a
caret). Let $p({\bf \hat{w}} | \rmsub{w}{\ti{0}})$ denote the pdf of
${\bf \hat{w}}$, given the true value of $\rmsub{w}{\ti{0}}$. One defines
${\bf \hat{w}}$ to be {\em unbiased \/} if the expected value of
${\bf \hat{w}}$ is equal to $\rmsub{w}{\ti{0}}$. In general the bias, $B$, of
${\bf \hat{w}}$ at true log distance $\rmsub{w}{\ti{0}}$ is given by:-
\begin{equation}
\label{eq:biasdefn}
\centering
  B({\bf \hat{w}},\rmsub{w}{\ti{0}}) \quad = \quad
  \int {\bf \hat{w}} p({\bf \hat{w}} | \rmsub{w}{\ti{0}}) d {\bf \hat{w}}
  \quad - \quad \rmsub{w}{\ti{0}}
\end{equation}
Another important quantity which one can introduce is the
{\em mean square error \/} or {\em risk \/}, $R$, of an estimator,
defined by:-
\begin{equation}
\label{eq:riskdefn}
\centering
  R({\bf \hat{w}},\rmsub{w}{\ti{0}}) \quad = \quad
  \int ({\bf \hat{w}} - \rmsub{w}{\ti{0}}) ^{2} \, p({\bf \hat{w}} |
  \rmsub{w}{\ti{0}}) d {\bf \hat{w}}
\end{equation}
(c.f. eqs. (16) and (17) of HS). Note that for an unbiased estimator,
the risk is identically equal to the variance. Note also that both the bias
and risk are in general functions of the true log distance,
$\rmsub{w}{\ti{0}}$.
This fact indicates the essential difficulty of completely removing Malmquist
bias from galaxy distance estimators: the magnitude of the bias for any given
galaxy in general depends upon its true distance, which is unknown.

The definition of the bias of an estimator given by equation
\ref{eq:biasdefn} differs from that adopted in those treatments of
Malmquist bias which we may categorise as Bayesian -- most notably the
derivation of `Malmquist corrections' in \cite{lynden88} (hereafter LB) and
\cite{landsz92} (hereafter LS). In the Bayesian picture one regards the true
log distance of the sampled galaxies itself as a random variable, to which
one can ascribe some prior probability distribution,
$p({\bf{\rmsub{w}{\ti{0}}}})$,
based upon an assumed spatial density distribution and selection function.
(note that ${\bf{\rmsub{w}{\ti{0}}}}$ is now written in bold face).
Following the measurement of the log distance estimator, ${\bf \hat{w}}$,
for each galaxy one can define a posterior distribution,
$p({\bf{\rmsub{w}{\ti{0}}}}|{\bf \hat{w}})$, for ${\bf{\rmsub{w}{\ti{0}}}}$
conditional
upon ${\bf \hat{w}}$ which will differ from the prior. It is the properties of
this posterior distribution which LB and LS consider in defining an estimator
as unbiased. By applying Bayes' theorem one can derive an expression for
$p({\bf{\rmsub{w}{\ti{0}}}}|{\bf \hat{w}})$, viz:-
\begin{equation}
\label{eq:bayes}
\centering
  p({\bf{\rmsub{w}{\ti{0}}}}|{\bf \hat{w}}) \quad = \quad
  \frac{p({\bf \hat{w}}|{\bf{\rmsub{w}{\ti{0}}}})
  p({\bf{\rmsub{w}{\ti{0}}}})}
  {\int p({\bf \hat{w}}|{\bf{\rmsub{w}{\ti{0}}}})
  p({\bf{\rmsub{w}{\ti{0}}}})
  d {\bf{\rmsub{w}{\ti{0}}}}}
\end{equation}
where the likelihood function, $p({\bf \hat{w}}|{\bf{\rmsub{w}{\ti{0}}}})$, is
simply the pdf of ${\bf \hat{w}}$ conditional on true log distance
${\bf{\rmsub{w}{\ti{0}}}}$, as in equation \ref{eq:biasdefn} above.

LB and LS define ${\bf \hat{w}}$ as unbiased if the expected value of
${\bf{\rmsub{w}{\ti{0}}}}$ with respect to the posterior distribution,
$p({\bf{\rmsub{w}{\ti{0}}}}|{\bf \hat{w}})$, is equal to ${\bf \hat{w}}$. In
general the bias of ${\bf \hat{w}}$ is defined by:-
\begin{equation}
\label{eq:biasLSdefn}
\centering
  B({\bf \hat{w}},\rmsub{w}{\ti{0}}) \quad = \quad
  \int {\bf{\rmsub{w}{\ti{0}}}} p({\bf{\rmsub{w}{\ti{0}}}}|{\bf \hat{w}})
  d {\bf{\rmsub{w}{\ti{0}}}}
  \quad - \quad {\bf \hat{w}}
\end{equation}
By assuming a posterior distribution and likelihood function LB and LS
derive a Malmquist correction to remove the bias of their `raw' log
distance estimator (which they denote by $\rmsub{\em l \/}{e}$), so that the
corrected estimator is unbiased.

The question of which approach one should take to the definition (to say
nothing of the elimination!) of Malmquist bias is far from clear-cut, and
depends strongly upon the context in which galaxy distance estimators are
being used. For example \cite{dekeletal90} develop their {\sc potent}
distance error analysis from the Bayesian viewpoint and apply Malmquist
corrections to their raw distance estimates. They argue that this approach is
essential to their analysis due to the nature of the smoothing procedures
carried out in {\sc potent}.

We study the effects of biased distance estimators on
{\sc potent} in \cite{newsametal93}, and give a full
discussion of the broader statistical issues relating to the merits of the
frequentist and Bayesian descriptions elsewhere (c.f. \cite{simmons93},
\cite{hendryetal93}).
Although we concentrate on the frequentist description of Malmquist bias for
the remainder of this paper, our results nevertheless have crucial implications
for the Bayesian approach. This is because the Malmquist corrections defined
in LB and LS are derived on the assumption of a raw log distance estimator,
$\rmsub{\em l \/}{e}$, for each galaxy which is normally distributed with mean
value equal to the true log distance. If this condition is not met then the
Malmquist corrections derived from $\rmsub{\em l \/}{e}$ will {\em not}
eliminate Malmquist bias \cite{newsamparis93}.

In short, then, LB and LS make the crucial assumption that
$\rmsub{\em l \/}{e}$ is unbiased in the frequentist sense, in order to
define a corrected estimator which is unbiased in the Bayesian sense. It is
this fact which makes our discussion of (frequentist) unbiased estimators in
this paper extremely important for both frequentist {\em and} Bayesian
approaches to Malmquist bias.

Rather than assuming a pdf for our log distance estimator as in LB and LS, in
section \ref{sec:MPobs} we now derive the pdf in terms of the intrinsic
joint distribution of absolute magnitude and line widths, and the
observational selection effects.

\subsection{{\sc{the observed distribution of m and p}}}
\label{sec:MPobs}

Let the absolute magnitude, {\bf M}, and spatial position, {\bf \underline{r}},
of a galaxy be random variables.
Suppose we now introduce a third random variable, {\bf P}, which denotes some
intrinsic physical characteristic of the galaxy such that the measured value
of {\bf P} in general provides information on the value of {\bf M} -- i.e.
{\bf M} and {\bf P} are correlated. It is
convenient to identify {\bf P} explicitly as log line width, as we have been
doing up until now, although one should bear in mind that the
formalism holds more generally for {\em any \/} suitably correlated physical
variable.

Suppose next that neither {\bf M} nor {\bf P}
is correlated with {\bf \underline{r}}, so that we may meaningfully introduce
$\Psi ({\bf M},{\bf P})$, the intrinsic joint distribution of {\bf M} and
{\bf P}, which is independent of spatial position. Let
$N ({\bf M},{\bf P},{\bf \underline{r}}) d{\bf M} d{\bf P} dV$
denote the actual number of galaxies in volume element $dV$ at spatial
position {\bf \underline{r}} with absolute magnitude in the range
{\bf M} to ${\bf M} + d{\bf M}$ and log line width in the range
{\bf P} to ${\bf P} + d{\bf P}$. It then follows that:-
\begin{equation}
\label{eq:numgals}
\centering
  N ({\bf M},{\bf P},{\bf \underline{r}}) d{\bf M} d{\bf P} dV =
  \Psi ({\bf M},{\bf P}) n({\bf \underline{r}}) d{\bf M} d{\bf P} dV
\end{equation}
where $n({\bf \underline{r}})$ is the number density of galaxies at
{\bf \underline{r}}.

Consider now the joint distribution,
$\rho({\bf M},{\bf P},{\bf \underline{r}})$, of {\bf M}, {\bf P} and
{\bf \underline{r}}, for {\em observable \/} galaxies in a sample subject
to observational selection effects.
We characterise the selection effects by a selection
function, $S({\bf M},{\bf P},{\bf \underline{r}})$, defined as the
probability that a galaxy of absolute magnitude, {\bf M}, and log line width,
{\bf P}, at spatial position, {\bf \underline{r}}, would be observable.

An expression for $\rho({\bf M},{\bf P},{\bf \underline{r}})$
in terms of
$\Psi ({\bf M},{\bf P}) n({\bf \underline{r}}) d{\bf M} d{\bf P} dV$ and
$n({\bf \underline{r}})$ now follows easily:-
\begin{equation}
\label{eq:rhoMPr}
\centering
  \rho({\bf M},{\bf P},{\bf \underline{r}}) =
  \frac {\Psi ({\bf M},{\bf P}) n({\bf \underline{r}})
  S({\bf M},{\bf P},{\bf \underline{r}})}
  {\int\!\!\int\!\!\int \Psi ({\bf M},{\bf P}) n({\bf \underline{r}})
  S({\bf M},{\bf P},{\bf \underline{r}}) d{\bf M} d{\bf P} dV }
\end{equation}
Note that the selection function, $S({\bf M},{\bf P},{\bf \underline{r}})$,
does not measure the probability that a galaxy would actually be
observed: clearly this would depend on the true local number density
of galaxies, $n({\bf \underline{r}})$, which will in general be unknown.
$S({\bf M},{\bf P},{\bf \underline{r}})$ as defined here will be independent
of $n({\bf \underline{r}})$ and, moreover, will also
be independent of direction provided that one has corrected for the directional
dependence of galactic extinction. A number of standard observational methods
exist for carrying out these corrections (c.f. \cite{sandtamm81},
\cite{burheil82}).

Because $S({\bf M},{\bf P},{\bf \underline{r}})$ is defined independently of
direction it is meaningful to consider the distribution,
$\phi({\bf M},{\bf P} | r_{0})$, of absolute magnitude and log
line width for observable galaxies conditional on true distance, $r_{0}$, or
equivalently on true log distance, $w_{0}$. It follows from equation
\ref{eq:rhoMPr} that $\phi({\bf M},{\bf P} | w_{0})$ is given by:-
\begin{equation}
\label{eq:phiMP_r}
\centering
  \phi({\bf M},{\bf P} | w_{0}) =
  \frac {\Psi ({\bf M},{\bf P}) S({\bf M},{\bf P}, w_{0})}
  {\int\!\!\int \Psi ({\bf M},{\bf P})
  S({\bf M},{\bf P}, w_{0}) d{\bf M} d{\bf P} }
\end{equation}
Note that this distribution is independent of the local number density,
$n({\bf \underline{r}})$, of galaxies. Although this useful property of
conditional distributions was pointed out by \cite{neyman52}, it would seem
that its relevance to Tully-Fisher type relations has not been widely
appreciated. The joint distribution of ${\bf M}$ and ${\bf P}$ at given
distance is generally derived on the assumption of a uniform spatial number
density (c.f. \cite{teerikorpi84}, \cite{bicknell92}). We see from equation
\ref{eq:phiMP_r} that such an assumption is in fact unnecessary, and in
particular $\phi({\bf M},{\bf P} | w_{0})$ is identical for both field and
cluster galaxies -- provided of course that one can assume the intrinsic
joint distribution $\Psi ({\bf M},{\bf P})$ to be independent of environment.

\subsection{{\sc{bias of general linear estimator}}}
\label{sec:biasgenlin}

Ignoring absorption and cosmological effects, the following equation
relates the apparent and absolute magnitudes of a galaxy at given true log
distance, $\rmsub{w}{\ti{0}}$:-
\begin{equation}
\label{eq:appabsM}
\centering
  {\bf m} = {\bf M} + 5 \rmsub{w}{\ti{0}} + \kappa
\end{equation}
Here $\kappa$ is a constant which depends
upon our units of distance. e.g.
if distances are measured in Mpc then $\kappa = 25$. If distances are
measured in kms$^{-1}$ by tying the calibration of one's distance estimator
to a cluster at some assumed redshift distance (as is commonly the case in
the literature) then $\kappa = 15 - 5 \log h$.

A sensible form for a general linear estimator,
${\bf{\rmsub{\hat{w}}{\ti{GL}}}}$,
of $\rmsub{w}{\ti{0}}$ is now clearly given by:-
\begin{equation}
\label{eq:whatgl}
\centering
  {\bf{\rmsub{\hat{w}}{\ti{GL}}}} \quad = \quad 0.2
  ( {\bf m} - {\bf \hat{M}} - \kappa) \quad = \quad
  0.2 ( {\bf m} - a {\bf P} - b - \kappa)
\end{equation}
where ${\bf \hat{M}} = a {\bf P} + b$, and $a$ and $b$ are constants.
(c.f. eq. \ref{eq:TFreln} above).

By combining equations \ref{eq:phiMP_r}, \ref{eq:appabsM} and \ref{eq:whatgl}
we can determine the joint distribution function of ${\bf m}$ and ${\bf P}$
for observable galaxies -- and from that the pdf of
${\bf{\rmsub{\hat{w}}{\ti{GL}}}}$ -- conditional on $\rmsub{w}{\ti{0}}$.
There is a somewhat more direct route to the same result, however.
Substituting equation \ref{eq:appabsM} back into equation \ref{eq:whatgl} and
rearranging we obtain:-
\begin{equation}
\label{eq:whatglc}
\centering
  {\bf{\rmsub{\hat{w}}{\ti{GL}}}} - \rmsub{w}{\ti{0}} =
  0.2 \, ( {\bf M} - a {\bf P} - b)
\end{equation}
Equation \ref{eq:whatglc} is of little practical use in defining
${\bf{\rmsub{\hat{w}}{\ti{GL}}}}$ since both ${\bf M}$ and
$\rmsub{w}{\ti{0}}$ are unknown. However an expression for the bias of
${\bf{\rmsub{\hat{w}}{\ti{GL}}}}$ now follows directly, viz:-
\begin{equation}
\label{eq:biasdefnb}
\centering
  B({\bf{\rmsub{\hat{w}}{\ti{GL}}}},\rmsub{w}{\ti{0}}) =
  0.2 \, (\, E({\bf M} | \rmsub{w}{\ti{0}}) \, - \, a
  \, E({\bf P} | \rmsub{w}{\ti{0}}) \, - b)
\end{equation}
where the expected values of $\bf{M}$ and $\bf{P}$ are with respect to the
joint distribution function for observable galaxies given by equation
\ref{eq:phiMP_r}.

Equation \ref{eq:biasdefnb} is valid for a completely general
selection function of $\bf{M}$, $\bf{P}$ and $\rmsub{w}{\ti{0}}$.
Consequently, the
expected values of $\bf{M}$ and $\bf{P}$ are both, in general, functions of
$\rmsub{w}{\ti{0}}$ and it is this fact which makes the complete elimination of
Malmquist bias from a linear estimator impossible in the general case: one
cannot choose values of the constants a and b which define the distance
estimator so that equation \ref{eq:biasdefnb} is identically zero
{\em for all \/} true distances.
To make any further progress towards identifying an unbiased distance
estimator requires making some assumptions about the nature of the
distribution function, $\phi({\bf M},{\bf P} | \rmsub{w}{\ti{0}})$.

\subsection{{\sc{schechter's solution for an unbiased estimator}}}
\label{sec:schech}

We can rewrite the intrinsic joint distribution function,
$\Psi ({\bf M},{\bf P})$, of ${\bf M}$ and ${\bf P}$ as follows:-
\begin{equation}
\label{eq:psifact}
\centering
  \Psi ({\bf M},{\bf P}) \quad = \quad \Psi ({\bf M})
  \Psi ({\bf P} | {\bf M})
\end{equation}
$\Psi ({\bf M})$ is just the galaxy luminosity function, well described by
e.g. a Schechter function or a gaussian, but regarded as an arbitrary
function for the moment. Note that this factorisation does {\em not \/} require
any assumption about in which variable lies the scatter in the Tully-Fisher
relation, but is valid in the completely general (and more realistic!) case
of scatter in both variables.
Taking as our lead the approach of \cite{schechter80}, suppose we now make
the following two crucial assumptions:-

\begin{enumerate}
\item{the selection function is {\em independent \/} of ${\bf P}$}
\item{the conditional
expectation of {\bf P} given {\bf M} is {\em linear \/} in {\bf M}, i.e.:-}
\end{enumerate}
\begin{equation}
\label{eq:condpm}
\centering
  E({\bf P} | {\bf M} ) = \alpha {\bf M} + \beta
\end{equation}
where $\alpha$ and $\beta$ are constants, equal to the slope and zero point
of the regression line of ${\bf P}$ upon ${\bf M}$.
With these two assumptions equation \ref{eq:biasdefnb} reduces to:-
\begin{equation}
\label{eq:biasdefnsch}
\centering
  B({\bf{\rmsub{\hat{w}}{\ti{GL}}}},\rmsub{w}{\ti{0}}) \quad = \quad
  0.2 \, ( \, (1 - \alpha a ) E({\bf M} | \rmsub{w}{\ti{0}}) \, -
  \, b \, - \, a \beta \, )
\end{equation}
from which one sees that if $a = \alpha^{-1}$ and
$b = - \beta \alpha^{-1}$ in equation \ref{eq:biasdefnsch}, then the
bias of ${\bf{\rmsub{\hat{w}}{\ti{GL}}}}$ is zero {\em for all \/} values of
$\rmsub{w}{\ti{0}}$. In other words this solution identifies an unbiased log
distance estimator, ${\bf{\rmsub{\hat{w}}{\ti{I}}}}$, viz:-
\begin{equation}
\label{eq:schdefn}
\centering
  {\bf{\rmsub{\hat{w}}{\ti{I}}}} = 0.2 \, ( {\bf m} - \alpha^{-1}
  ( {\bf P} - \beta) \, - \, \kappa)
\end{equation}
We use the subscript `I' since this unbiased solution corresponds exactly to
estimator one obtains from applying the inverse Tully-Fisher relation --
i.e. regressing line widths on magnitudes -- in complete concordance with
Schechter's result.

To fix these ideas with a specific example, consider again the case where
${\bf M}$ and ${\bf P}$ are jointly normally distributed. This case certainly
satisfies the assumption that the conditional expectation of {\bf P} given
{\bf M} be linear in {\bf M}. Comparing equations \ref{eq:TFinv} and
\ref{eq:condpm} we see that $\alpha = \rho
\frac{\rmsub{\sigma}{\ti{P}}}{\rmsub{\sigma}{\ti{M}}}$ and
$\beta = \rmsub{P}{\ti{0}} - \rho
\frac{\rmsub{\sigma}{\ti{P}}}{\rmsub{\sigma}{\ti{M}}} \rmsub{M}{\ti{0}}$,
which implies the following expression for the unbiased `inverse' estimator:-
\begin{equation}
\label{eq:whatsch}
\centering
  {\bf{\rmsub{\hat{w}}{\ti{I}}}} \quad = \quad
  0.2 \, ( {\bf m} \, - \, \rmsub{M}{\ti{0}} \, - \,
  \frac{\rmsub{\sigma}{\ti{M}}}{\rho \rmsub{\sigma}{\ti{P}}}
  ({\bf P} - \rmsub{P}{\ti{0}}) \, - \, \kappa )
\end{equation}
It is instructive to compare ${\bf{\rmsub{\hat{w}}{\ti{I}}}}$ with the
`direct'
log distance estimator, ${\bf{\rmsub{\hat{w}}{\ti{D}}}}$, corresponding to the
direct regression of magnitudes on line widths. The values of the constants
$a$ and $b$ for this case follow from equation \ref{eq:TFdir}, and give:-
\begin{equation}
\label{eq:whatdir}
\centering
  {\bf{\rmsub{\hat{w}}{\ti{D}}}} \quad = \quad
  0.2 \, ( {\bf m} \, - \, \rmsub{M}{\ti{0}} \, - \,
  \frac{\rho \rmsub{\sigma}{\ti{M}}}{\rmsub{\sigma}{\ti{P}}}
  ({\bf P} - \rmsub{P}{\ti{0}}) \, - \, \kappa )
\end{equation}
which differs from equation \ref{eq:whatsch} only in the
switching of the correlation coefficient, $\rho$, from denominator to
numerator, reflecting the different slope of the direct regression line
(c.f. Figures (1) and (2).
The bias of ${\bf{\rmsub{\hat{w}}{\ti{D}}}}$ now follows from equation
\ref{eq:biasdefnsch}, after a little reduction:-
\begin{equation}
\label{eq:biasdir}
\centering
  B({\bf{\rmsub{\hat{w}}{\ti{D}}}},\rmsub{w}{\ti{0}}) \quad = \quad
  0.2 \, (1 - \rho^{2} ) \{ E({\bf M} | \rmsub{w}{\ti{0}}) \, - \,
  \rmsub{M}{\ti{0}} \}
\end{equation}
Several points emerge from this equation. Firstly notice that when $\rho = 0$,
i.e. when ${\bf P}$ and ${\bf M}$ are uncorrelated, then the bias of
${\bf{\rmsub{\hat{w}}{\ti{D}}}}$ reduces to the bias of the `naive' estimator,
${\bf{\rmsub{\hat{w}}{\ti{n}}}}$, of log distance defined in HS and
\cite{hendry92} by:-
\begin{equation}
\label{eq:whatnaive}
\centering
  {\bf{\rmsub{\hat{w}}{\ti{n}}}} \quad = \quad 0.2
  ( {\bf m} - \rmsub{M}{\ti{0}} - \kappa)
\end{equation}
i.e. assuming that all galaxies are standard candles of absolute magnitude,
$\rmsub{M}{\ti{0}}$, and ignoring the effects of Malmquist bias. This is not
surprising, since when $\rho = 0$ the measured log line width provides no
additional information about the value of ${\bf M}$. The second point to note
is that as $| \rho |$ tends to unity, on the other hand, the bias of
${\bf{\rmsub{\hat{w}}{\ti{D}}}}$ tends to zero at all true distances. Again
this
follows automatically from the fact that as $| \rho | \rightarrow 1$ the
direct and inverse regression lines become collinear, and
${\bf{\rmsub{\hat{w}}{\ti{D}}}}$ and ${\bf{\rmsub{\hat{w}}{\ti{I}}}}$ are
identical.
Lastly note that if there are {\em no \/} magnitude selection effects then
${\bf{\rmsub{\hat{w}}{\ti{D}}}}$ is again unbiased at all true distances,
simply because we then have $E({\bf M} | \rmsub{w}{\ti{0}}) =
\rmsub{M}{\ti{0}}$ for all
$\rmsub{w}{\ti{0}}$. It is easy to see that this result is true for an
arbitrary joint intrinsic distribution function, $\Psi ({\bf M},{\bf P})$, in
the absence of selection effects.

Finally we consider the risk and the higher moments of the
${\bf{\rmsub{\hat{w}}{\ti{GL}}}}$ distribution. We can do this most easily by
introducing a new random variable ${\bf t} = {\bf P} - (\alpha {\bf M} +
\beta)$. This allows us to rewrite equation \ref{eq:whatglc} as follows:-
\begin{equation}
\label{eq:whatglt}
\centering
  {\bf{\rmsub{\hat{w}}{\ti{GL}}}} - \rmsub{w}{\ti{0}} =
  0.2 \, [ (1 - \alpha A) {\bf M} \, - \, B \, - \, \beta A \, - \,
  A {\bf t} ] )
\end{equation}
For the unbiased inverse estimator we see that all but the final term
of the right hand side vanishes. It follows immediately from this that the
moments of ${\bf{\rmsub{\hat{w}}{\ti{I}}}} - \rmsub{w}{\ti{0}}$ are equal
simply to a constant multiple of the moments of, ${\bf t}$,
{\em independent of the true log distance! \/}. Moreover, since we are
assuming that $E({\bf P} | {\bf M} ) = \alpha {\bf M} + \beta$, it follows
that the probability distribution of ${\bf{\rmsub{\hat{w}}{\ti{I}}}}$ is
identical in shape to the intrinsic conditional distribution,
$\Psi ({\bf P} | {\bf M})$. This latter distribution is generally modelled to
be gaussian (c.f. \cite{teerikorpi84}, \cite{bicknell92}), thus implying
that the inverse estimator is normally distributed, unbiased, and of constant
variance at all true log distances.

As we recalled in section \ref{sec:what}, these are precisely the properties
assumed for the raw log distance estimator, $\rmsub{\em l \/}{e}$, in LB and
LS. Our results confirm, therefore, that ${\bf{\rmsub{\hat{w}}{\ti{I}}}}$ is
the correct raw log distance estimator to use in defining Malmquist
corrections.

It follows from equation \ref{eq:whatglt}, on the other hand, that
${\bf{\rmsub{\hat{w}}{\ti{D}}}}$ will {\em not} be normally distributed for
all $\rmsub{w}{\ti{0}}$, and in fact will lead to incorrect Malmquist
corrections if these are derived on the assumption of a normal raw estimator.
Notwithstanding this important result, to our knowledge a direct linear
regression has been used exclusively to date in the derivation both
homogeneous and inhomogeneous Malmquist corrections in the literature (c.f.
\cite{lynden88}, \cite{bertetal90}, \cite{potiras93}, \cite{courteauetal93}).
We examine the consequences of this incorrect choice of raw distance
estimator in \cite{newsametal93} and \cite{newsamparis93}.

\subsection{{\sc{properties of the unbiased `inverse' estimator}}}
\label{sec:props}

It is instructive to summarise the properties of the inverse estimator,
${\bf{\rmsub{\hat{w}}{\ti{I}}}}$, which we have thus far confirmed or
determined, and add several further results which follow easily from them.

\begin{enumerate}
\item{In a sample subject to observational selection effects, provided that
the measurements of line width are selection-free and the conditional
expectation of line width at given absolute magnitude is linear in {\bf M},
then it is possible to define a general linear estimator of log distance
which is unbiased at all true distances, and the appropriate linear combination
corresponds exactly to the estimator derived from a regression of (log) line
widths upon magnitudes, as prescribed in \cite{schechter80}. This result is
valid in the general case where one accounts for intrinsic and observational
scatter in both variables, and does not require the assumption that the
scatter lies only in line widths.}

\item{The `inverse' estimator thus defined is the only unbiased linear
estimator of log distance. Any other linear combination of magnitude and
log line width, and in particular any other regression line, yields an
estimator which is biased at all true distances for a magnitude selection
function. Examples
of biased regression lines in this case include, therefore, not only the
direct regression used by e.g. \cite{lynden88} (in its equivalent form
for the \dns relation), but also the orthogonal regression (accounting for
residuals on both observables -- c.f. \cite{giraud87}); `bisector' regression
(i.e. the line which bisects the direct and inverse regression lines -- c.f.
\cite{ptully88}) and mean (i.e. the line whose slope is the arithmetic mean
of the direct and inverse lines -- c.f. \cite{mouldetal93}) regression lines.}

\item{The shape of the pdf, and hence in particular the risk (or
equivalently variance), of the inverse estimator is constant at all true
distances. It follows from this property that confidence intervals derived
from the inverse estimator, following the method outlined in HS, are of
constant width. For any other general linear estimator, on the other hand, the
shape of the pdf is severely distorted at large true distances as luminosity
selection effects become significant.}

\item{The pdf of the inverse estimator is, in fact, identical in shape to the
intrinsic pdf of log line width conditional upon absolute magnitude. If the
latter distribution is normal and of constant variance, as is commonly assumed,
then so too will be the pdf of the inverse estimator. It is therefore the
correct choice of `raw' log distance estimator for the derivation of Malmquist
corrections.}

\item{The unbiased property of the inverse estimator is true for an arbitrary
luminosity function and magnitude selection function, and is independent of the
true number density distribution of galaxies. This is a particularly useful
property, since it follows from equations \ref{eq:phiMP_r} and
\ref{eq:biasdefnb} that the bias of any other linear estimator will depend
explicitly upon the form of the luminosity function and magnitude selection
effects, so that any attempt to correct for or reduce the bias would
necessarily be model dependent. Indeed, \cite{bicknell92} shows that the
magnitude of the bias of the direct regression line is substantially
different for gaussian and Schechter luminosity functions.}

\item{One may also define an unbiased log distance estimator for other distance
indicators, including the \dns and magnitude-colour relations, subject to the
same condition that there be one observable free from selection, but
{\em not \/} requiring one observable to be distance-independent. In a
diameter-complete survey, for example, one may construct an unbiased distance
estimator from the observed angular diameter and apparent magnitude. As above,
it is straightforward to show that this unbiased estimator corresponds
exactly to the regression of the selection-free observable upon the other
observable.}

\item{The inverse estimator is an unbiased estimator of log
distance: consequently the corresponding distance estimator is biased. It is
a simple matter, however, to define a corresponding unbiased distance
estimator, particularly in the case where ${\bf{\rmsub{\hat{w}}{\ti{I}}}}$
is normally distributed (c.f. \cite{lynden88}).}
\end{enumerate}

\subsection{{\sc{unbiased estimators in more realistic cases}}}
\label{sec:realistic}

Although we have striven to show in this paper that the definition of unbiased
estimators
following the prescription of \cite{schechter80} rests on few assumptions and
is otherwise a very general result, one must nevertheless accept that even
these modest assumptions may not be met in most practical situations. In
particular, if neither observable is free from selection effects then an
unbiased estimator formally {\em cannot \/} be defined as a simple linear
combination of the observables.

In the context of both the Tully-Fisher and \dns relations,
however, the problem of ${\bf P}$ selection is somewhat less important than
one might expect. Most surveys will be subject to a lower selection limit on
line width or velocity dispersion: e.g. it will not be possible to measure
accurately velocity dispersions of the order of 150kms$^{-1}$ \cite{lynden88}.
The interesting -- and very useful -- property of this selection limit is,
however, that it becomes increasingly
{\em less \/} important at larger distances. This is easy to understand,
since at large distances only intrinsically brighter (or larger), and thus
sufficiently large line width, galaxies will be observable. In other words,
at large distances the galaxies which are `lost' to the survey due to their
small velocity widths would have been unobservable in any case, owing to
their faint luminosity.

As an illustrative example, Figure (3) shows the bias of the
inverse log distance estimator derived from the combined Virgo and Ursa
Major calibrating sample of \cite{ptully88}, and assuming a sharp I-band
magnitude limit at $\rmsub{I}{\ti{LIM}} = 14$. The curves show the bias of
${\bf{\rmsub{\hat{w}}{\ti{I}}}}$ as a function of true distance (expressed in
kms$^{-1}$) for three different line width selection limits. Note that
the effect of ${\bf P}$ selection is to introduce a positive bias -- in
contrast to the negative Malmquist bias caused by an upper limit on
observable apparent magnitude. The effect is clearly very small, however.
A bias of 0.01 in ${\bf{\rmsub{\hat{w}}{\ti{I}}}}$ corresponds to a systematic
distance error of $\sim 2 \%$. Hence, one sees that the effect of a
line width limit as large as $\rmsub{P}{\ti{LIM}} = 200$ kms$^{-1}$ is
negligible,
and even with a limit of $\rmsub{P}{\ti{LIM}} = 250$ kms$^{-1}$ the effect
can still be ignored at cosmologically interesting distances in this case.

In the event that selection effects on ${\bf P}$ are large enough to be
significant -- or, for example, if the line width selection cannot be well
described by a sharp limit, independent of distance and morphological type --
one can adopt an iterative method to reduce Malmquist bias --
although such an approach will necessarily be model dependent. We discussed
this method in HS, for the case of an estimator which is a function of
apparent magnitude only -- so that Schechter's ideas are inapplicable. The
extension to estimators of Tully-Fisher type is straightforward, however.
Let ${\bf{\hat{w}}}({\bf m},{\bf P})$ denote an estimator of log distance as
before. Rearranging equation \ref{eq:biasdefn} observe that we may write:-
\begin{equation}
\label{eq:biasiter}
\centering
  E({\bf \hat{w}} ({\bf m},{\bf P}) | \rmsub{w}{\ti{0}}) \quad = \quad
  \rmsub{w}{\ti{0}} \quad + \quad
  B({\bf \hat{w}},\rmsub{w}{\ti{0}})
\end{equation}
This is essentially equation (19) of HS, in the equivalent form for an
estimator of log distance.

Although we cannot use equation \ref{eq:biasiter} to remove the bias of
${\bf{\hat{w}}}({\bf m},{\bf P})$
exactly, since the true log distance $\rmsub{w}{\ti{0}}$ is unknown, suppose
we form a new estimator, ${\bf{\rmsub{\hat{w}}{\ti{1}}}}
({\bf m},{\bf P})$, defined by:-
\begin{equation}
\label{eq:estiter}
\centering
  {\bf{\rmsub{\hat{w}}{\ti{1}}}}({\bf m},{\bf P}) \quad = \quad
  {\bf{\hat{w}}}({\bf m},{\bf P}) \quad - \quad
  B({\bf \hat{w}}({\bf m},{\bf P}),\rmsub{w}{\ti{0}} =
  {\bf \hat{w}}({\bf m},{\bf P}))
\end{equation}
In other words for each ${\bf m}$ and ${\bf P}$ we subtract from
${\bf{\hat{w}}}({\bf m},{\bf P})$ the bias of the estimator assuming that
the true log distance is equal to its estimated value. (c.f.
eq. (20) of HS). One can then compute the bias of the new estimator,
${\bf{\rmsub{\hat{w}}{\ti{1}}}}({\bf m},{\bf P})$, apply equation
\ref{eq:estiter} again to define ${\bf{\rmsub{\hat{w}}{\ti{2}}}}
({\bf m},{\bf P})$ in terms of ${\bf{\rmsub{\hat{w}}{\ti{1}}}}$, and so on.

It is not obvious that the above iterative scheme will in all cases converge
to an unbiased estimator. In fact we have shown \cite{hendry92} that this
is not the case for estimators which are functions of apparent magnitude only.
Numerical studies indicate that convergence is achieved for the Tully-Fisher
case with selection on both observables, however, provided that the scatter
in the intrinsic joint distribution of ${\bf M}$ and ${\bf P}$ is not too
large.

Perhaps a more serious problem in defining unbiased distance estimators lies
in the calibration of the
distance relation itself. In order to define the inverse estimator (or indeed
the direct estimator), one must determine the parameters of the
joint distribution of ${\bf M}$ and ${\bf P}$ -- e.g. the five parameters
$\rmsub{M}{\ti{0}}$, $\rmsub{P}{\ti{0}}$, $\rmsub{\sigma}{\ti{M}}$,
$\rmsub{\sigma}{\ti{P}}$ and
$\rho$ in the bivariate normal case. It is obviously of great importance,
therefore, to ensure that the estimates of these parameters obtained from one's
calibrating sample accurately reflect their true intrinsic values. It has
been suggested (c.f. \cite{tammann87}) that the scatter measured
in distance relations underestimates the true scatter -- leading one to
suppose a less serious contribution from Malmquist bias -- simply because the
number of calibrating galaxies is insufficient to accurately determine the
slope and zero point of the relation.

We have addressed this question in some quantitative detail, carrying out
numerical experiments on artificial cluster samples of a range of different
sizes and true parameters, in order to determine how many calibrators are
required to achieve a given level of accuracy in the fitted Tully-Fisher
slope. As an illustration, Figure (4) shows the results of
Monte Carlo simulations carried out assuming a bivariate normal model for the
distribution of ${\bf M}$ and ${\bf P}$ and adopting as true parameter
values those given by the Fornax cluster used in the calibration of the
Mathewson galaxy survey (c.f. \cite{mathewson92}). The bold and dotted lines
show the true inverse and direct regression line slopes respectively, while
the two curves show $1 \sigma$ confidence limits for the estimated inverse
regression line slope as a function of the number of galaxies in the
calibrating sample.

One can see from Figure (4) that for calibrating samples
containing less than $\sim 40$ galaxies, the dispersion of the estimated
slope of the inverse regression line is greater than the difference between
the true slopes of the inverse and direct regression lines.
Hence one requires a calibrating sample of over 40 galaxies in order that the
scatter in the slope of the inverse regression line due to sampling error be
smaller than the difference between the slopes of the two lines.

Putting this another way, with a considerably smaller sample of calibrators
there is a strong possibility that the bias in the
(supposedly unbiased!) inverse estimator due to incorrect determination of the
estimator slope will be {\em larger \/} than the Malmquist bias of the direct
estimator.

Clearly, then, it is important to use as large a calibrating sample as possible
to minimise this problem. One solution is to combine data from several
different clusters, as  in \cite{ptully88} and \cite{mouldetal93},
combining two samples from the Virgo and Ursa Major clusters, whose distance
moduli have been found to be equal. \cite{latthen93}) discuss two different
methods of tackling the problem of combining calibration
data from clusters at different distances, and obtaining optimal estimates of
the slope and zero point of the distance relation simultaneously with
relative distances to each cluster.

Of course another way in which the problems of sampling error can be reduced is
by identifying distance relations of intrinsically smaller scatter. In section
\ref{sec:genlin3} we consider how one might achieve this by defining
estimators which are functions of more than two observables.

\section{{\sc{estimators of distance using three or more observables}}}
\label{sec:genlin3}

In this section we briefly discuss the properties of distance estimators which
are defined as a function of apparent magnitude and two other observable
quantities, such as one might define in extending the Tully-Fisher relation
to include the observed angular diameter of spiral galaxies. Of particular
interest is the question of whether one may still define unbiased estimators
in this case, analogous to the P on M estimator of the previous section, and if
so whether one may construct unbiased estimators which have a smaller risk
than their two-variable counterparts.

One can carry out an analysis which follows closely the formulation adopted in
section \ref{sec:genlin}: i.e. first one derives the joint distribution at
given true log distance, $\rmsub{w}{\ti{0}}$, after
accounting for observational selection effects, of the random variables --
${\bf M}, {\bf P}$ and ${\bf D}$ say, denoting for example absolute magnitude,
log line width and log of absolute diameter -- in terms of their intrinsic
joint
distribution and selection function to obtain an expression analogous to
equation \ref{eq:phiMP_r}, viz:-
\begin{equation}
\label{eq:phiMPD_r}
\centering
  \phi({\bf M},{\bf P},{\bf D} | w_{0}) =
  \frac {\Psi ({\bf M},{\bf P},{\bf D}) S({\bf M},{\bf P},{\bf D}, w_{0})}
  {\int\!\!\int\!\!\int \Psi ({\bf M},{\bf P},{\bf D})
  S({\bf M},{\bf P},{\bf D}, w_{0}) d{\bf M} d{\bf P} d{\bf D} }
\end{equation}
One can then determine, for a general linear combination of the observables,
the distribution, bias and risk of this `general linear' estimator and, as
before, identify for which values the estimator is unbiased. The details of
these calculations are somewhat tedious and add little to the previous
analysis for two variables. We present, therefore, a summary of the main
results for the 3-variable case.

We considered two cases: firstly where only one of the three observables
is free from observational selection, and secondly where two observables are
selection-free. In both cases it was possible to define an unbiased estimator
of log distance by appropriate linear combination of the observables. The
values of the coefficients corresponding to the unbiased solution were given in
terms of the parameters of the intrinsic distribution function,
$\Psi ({\bf M},{\bf P},{\bf D})$, as in the two variable case. To take a
specific example, if ${\bf M},{\bf P} and {\bf D}$ were jointly normally
distributed, then the coefficients depend solely upon the mean values,
dispersions and correlation coefficients of the trivariate normal distribution.

In the first case where only one observable is selection-free, we
found that an unbiased estimator can, in general, be defined {\em only \/} as a
linear combination of all three observables. This has important consequences
for our earlier results. In the case of the Tully-Fisher relation, for
example, if one's sample is subject to both diameter and magnitude selection
then the inverse estimator defined in section \ref{sec:genlin} using only
apparent magnitude and log line width will no longer be unbiased. This
is because the selection on diameters affects the joint distribution of {\bf m}
and {\bf P}, since the galaxy diameter is correlated with these variables. A
similar effect is discussed in \cite{mouldetal93}, where selection on
diameter and surface brightness `pollutes' the distribution of {\bf m} and
{\bf P} and affects the bias of the Tully-Fisher relation. Clearly, therefore,
great care must be taken in ensuring no additional observables introduce
selection `by proxy' into one's samples. The fact that an observable does not
appear in the definition of one's distance estimator does {\em not \/} imply
that it can have no effect on the bias of that estimator.

In the second case, where two observables are free from selection, a
rather different picture emerges. Taking again the example of magnitude,
line width and diameter to fix ideas, we found that in this case the inverse
estimator defined in section \ref{sec:genlin} is still unbiased at all true
distances, so that Schechter's prescription is still be valid. The inverse
estimator is, however, no longer the {\em only \/} unbiased estimator of log
distance - although it is still the only unbiased estimator formed from a
linear combination of magnitude and line width alone.
By forming an estimator from three observables, we have sufficient
freedom to define an unbiased estimator of minimum variance, and one may
show that the variance of this optimal 3-variable estimator is always less
than or equal to that of the inverse estimator defined by magnitude and
line width alone.

The precise factor, $\Delta$, by which the addition of a third observable,
${\bf D}$, reduces the variance of the inverse estimator depends only upon the
values of the correlation coefficients between the three observables
(c.f. \cite{hendry92}). As an illustration, consider the specific case
where ${\bf M}$, ${\bf P}$ and ${\bf D}$ are jointly normally distributed,
with correlation coefficients denoted by $\rmsub{\rho}{\ti{MP}}$,
$\rmsub{\rho}{\ti{MD}}$, and $\rmsub{\rho}{\ti{PD}}$. In this case $\Delta$
is given by the following expression:-
\begin{equation}
\label{eq:deltarhos}
\centering
  \Delta =
  \frac{\rmsub{\rho}{\ti{MP}}^{2} [ \, 1 - \, ( {\rmsub{\rho}{\ti{MP}}^{2}} +
  {\rmsub{\rho}{\ti{MD}}^{2}} + {\rmsub{\rho}{\ti{PD}}^{2}} ) \, + \,
  2 {\rmsub{\rho}{\ti{MP}}} {\rmsub{\rho}{\ti{MD}}}
  {\rmsub{\rho}{\ti{PD}}} \, ] }
  { ( 1 - {\rmsub{\rho}{\ti{MP}}^{2}} ) [ {\rmsub{\rho}{\ti{MP}}^{2}} \, - \,
  2 {\rmsub{\rho}{\ti{MP}}} {\rmsub{\rho}{\ti{MD}}}
  {\rmsub{\rho}{\ti{PD}}} \, + \, {\rmsub{\rho}{\ti{MD}}^{2}} ] }
\end{equation}
Figures (5), (6) and (7) show
respectively scatter diagrams for the I-band magnitude versus log line width,
magnitude versus log diameter and log diameter versus log line width relations
for the Fornax cluster, determined from the Mathewson galaxy redshift survey.
It is clear from these figures that a very good correlation exists between
all three observables, and the correlation coefficients for this calibrating
sample were found to be $\rmsub{\rho}{\ti{MP}} = -0.985$,
$\rmsub{\rho}{\ti{MD}} = -0.963$ and $\rmsub{\rho}{\ti{PD}} = 0.942$.
Notwithstanding
the fact that the Fornax cluster is a rather small calibrating sample, in
the light of our remarks in section \ref{sec:realistic}, if we assume these
correlation coefficients to be equal to the intrinsic values for the
magnitude -- diameter -- line width relation then substituting in equation
\ref{eq:deltarhos} gives a value of $\Delta = 0.64$. In other words the
variance of the 3 variable estimator is more than $35 \%$ smaller than that of
the corresponding P on M estimator. This corresponds to a reduction in the mean
distance error dispersion from $\sim 20\%$ to around $15 \%$.

It would seem clear, therefore, that utilising the measurements of a third
observable can offer a means of signifcantly reducing the dispersion  of
unbiased distance estimators, and thus obtaining more reliable distance
estimates. When such an observable is available -- as is the case in the
above example of the magnitude -- diameter -- linewidth relation, its use
would seem to be strongly advised.

\section{{\sc{conclusions}}}
\label{sec:conc}

In this paper we have studied the properties of galaxy distance estimators
derived from combining measurements of two or more observables, as is the case
for the Tully-Fisher and \dns relations. We have considered the effects of
observational selection upon the distribution, bias and risk of these
estimators and have established that, subject to modest but crucial
assumptions, it is possible to define estimators which are {\em unbiased \/} at
all true distances, in confirmation of the results of \cite{schechter80}.
We have shown that these results are more general than is often
assumed in the literature: in particular, that one can define unbiased
distance estimators independently of the form of the magnitude selection
function and the local number density of galaxies, and almost independently of
the intrinsic joint distribution of magnitude and line width. Moreover, the
results are derived in the general case of observational and intrinsic
scatter on both correlated variables.

We have compared our treatment of Malmquist bias with other approaches which
have been adopted in the literature, and shown how the differences between
them can be understood as fundamentally different interpretations of the
nature of probability. Moreover, we have shown that when the distribution of
log line widths conditional on magnitudes is normal, then so too is the pdf of
the unbiased inverse estimator. It is therefore the only appropriate choice of
raw log distance estimator which is consistent with the assumptions made in
deriving homogeneous and inhomogeneous Malmquist corrections in the literature.

Finally, we have also considered how one can define unbiased estimators of
smaller variance by utilising additional, suitably correlated, observables.
In future work we will apply these multivariate estimators to the analysis of
real galaxy surveys, in order to extend and improve the optimal techniques
for smoothing and recovery of the peculiar velocity field described in
\cite{newsametal93} and \cite{simmonsetal93}.
\vspace{5mm}

{\sc Acknowledgments}

MAH acknowledges the use of the {\sc starlink} Microvax 3400 computer,
funded by a SERC grant to the Astronomy Centre at the University of
Sussex. During the course of this work MAH was supported by a
Carnegie Research Studentship at the University of Glasgow and a SERC
Research Fellowship at the University of Sussex.

\clearpage
\newpage

\noindent
{\bf{LIST OF FIGURES}}

\begin{description}
\item{ \bf{Figure 1.} Schematic Tully-Fisher relations, derived by applying a
direct and inverse linear regression to a complete calibrating sample -- e.g.
a nearby cluster.}
\item{ \bf{Figure 2.} The expected value of absolute magnitude conditional
upon log line width, and log line width conditional upon magnitude, in
a distribution subject to a sharp selection limit on absolute
magnitude -- e.g. a distant cluster. The shaded region
represents unobservable galaxies.}
\item{ \bf{Figure 3.} Bias of the inverse estimator with line width selection
effects, assuming a bivariate normal distribution for ${\bf M}$ and ${\bf P}$
with distribution parameters taken from the Virgo and Ursa Major composite
calibrating sample of \cite{ptully88}}
\item{ \bf{Figure 4.} $1 \sigma$ confidence limits for the sample estimate of
the slope of the inverse regression line as a function of the number of
galaxies in the calibrating sample. Distribution parameters are taken from
the Fornax cluster -- as determined in the Mathewson galaxy survey.}
\item{ \bf{Figure 5.} Scatter plot of the Tully-Fisher, I-band magnitude versus
log line width, relation for the Fornax cluster, derived from the
Mathewson redshift survey.}
\item{ \bf{Figure 6.} Scatter plot of the I-band magnitude versus
log diameter relation for the Fornax cluster, derived from the
Mathewson redshift survey.}
\item{ \bf{Figure 7.} Scatter plot of the log diameter versus
log line width relation for the Fornax cluster, derived from the
Mathewson redshift survey.}
\end{description}
\vfill
\end{document}